\documentclass[conference]{IEEEtran}
\IEEEoverridecommandlockouts
\usepackage{cite}
\usepackage{amsmath,amssymb,amsfonts}
\usepackage{algpseudocode}
\usepackage{graphicx}
\usepackage{textcomp}

\usepackage{xcolor}
\usepackage{bm}
\usepackage{psfrag}
\usepackage[utf8]{inputenc} 
\usepackage[T1]{fontenc}    
\usepackage{hyperref}       
\usepackage{url}            
\usepackage{booktabs}       
\usepackage{amsfonts}       
\usepackage{nicefrac}       
\usepackage{microtype}      
\usepackage{amsmath}
\usepackage[inline]{enumitem}
\DeclareGraphicsExtensions{.pdf,.jpeg,.png,  .eps}
\usepackage{psfrag}
\usepackage{epstopdf}
\usepackage{amsfonts}
\usepackage{mathrsfs}
\usepackage{pifont}
\usepackage{verbatim}
\usepackage{upgreek}
\usepackage{soul,color}
\usepackage{caption}
\usepackage{algorithm}
\usepackage{mdwmath}
\usepackage{mdwtab}
\usepackage{amssymb,amsmath}
\usepackage{amsmath}
\usepackage{amsthm}
\usepackage{epstopdf}
\usepackage{tikz}
\usepackage{scalefnt}
\usepackage{subfigure}
\usepackage{MnSymbol}
\usepackage{setspace}
\usepackage{mathtools}
\usepackage{psfrag}
\usepackage{cite}
\usepackage{xcolor, color, soul}
\sethlcolor{yellow}
\newcommand{\Diag}{\mbox{\boldmath\bf Diag}\, }

\label{English Chars}

\newcommand{\bH}{\textbf{H}}
\newcommand{\bh}{\textbf{h}}
\newcommand{\bI}{\textbf{I}}

\newcommand{\bR}{\textbf{R}}

\newcommand{\bW}{\textbf{W}}

\label{Roman Chars}

\newcommand{\bmu}{\mbox{\boldmath{$\mu$}}}

\label{Theorems}

\theoremstyle{remark}

\def\BibTeX{{\rm B\kern-.05em{\sc i\kern-.025em b}\kern-.08em
    T\kern-.1667em\lower.7ex\hbox{E}\kern-.125emX}}
\begin{document}

\title{MIMO Systems with Reconfigurable Antennas:\\ Joint Channel Estimation and Mode Selection
}

\author{\IEEEauthorblockN{Fariba Armandoust\IEEEauthorrefmark{1},
Ehsan Tohidi\IEEEauthorrefmark{1},  Martin~Kasparick\IEEEauthorrefmark{1},\\ Li Wang\IEEEauthorrefmark{2}, Ahmet Hasim Gokceoglu\IEEEauthorrefmark{2},  and
Slawomir Stanczak\IEEEauthorrefmark{1}}
\IEEEauthorblockA{\IEEEauthorrefmark{1}Fraunhofer Heinrich Hertz Institute,
Berlin, Germany,\\ \IEEEauthorrefmark{2}Huawei Technologies,
Stockholm, Sweden\\
Email: fariba.armandoust@hhi.fraunhofer.de, ehsan.tohidi@hhi.fraunhofer.de, martin.kasparick@hhi.fraunhofer.de}}


\maketitle

\begin{abstract}
   Reconfigurable antennas (RAs) are a promising technology to enhance the capacity and coverage of wireless communication systems. However, RA systems have two major challenges: $(i)$ High computational complexity of mode selection, and $(ii)$ High overhead of channel estimation for all modes. In this paper, we develop a low-complexity iterative mode selection algorithm for data transmission in an RA-MIMO system. Furthermore, we study channel estimation of an RA multi-user MIMO system. However, given the coherence time, it is challenging to estimate channels of all modes. We propose a mode selection scheme to select a subset of modes, train channels for the selected subset, and predict channels for the remaining modes. In addition, we propose a prediction scheme based on pattern correlation between modes. Representative simulation results demonstrate the system's channel estimation error and achievable sum-rate for various selected modes and different signal-to-noise ratios (SNRs).
\end{abstract}


\section{Introduction}
In modern communication systems, multiple input multiple output (MIMO) systems with multiple antennas at the transmitter and receiver are utilized to improve spectral efficiency without increasing the power or bandwidth requirements. Spatial diversity, spatial multiplexing, and beamforming are common approaches used in signal processing of MIMO systems to improve performance \cite{proakis2007digital, 9095425, 8537943}. 

Reconfigurable antennas (RAs) are a recent technology to support the increasing demand for high data rates and efficient spectrum utilization. RAs can alter operating frequencies, polarizations, and radiation patterns to accommodate changing operating requirements. Each reconfigurable mode of operation in RA-based systems is achieved by using technologies such as MEMS switching and semiconductor switching \cite{bahceci2016efficient,towfiq2018reconfigurable}. In this paper, the RA systems with radiation pattern reconfigurability are studied. The main idea of wireless systems with radiation pattern reconfigurability is to exploit the impact of different radiation patterns on the impulse response of the wireless channel \cite{bouida2015reconfigurable}. This antenna technology, combined with the MIMO systems, provides additional degrees of freedom to combat the adverse effect of the propagation channel and multi-user interference. Using RA elements and optimizing the system parameters, such as the physical structure of the channel, coding, and signal processing of MIMO systems can change the channel realization in favor of selected users. 

Using RA elements in MIMO systems leads to a channel estimation overhead since each antenna mode creates a different radiation pattern that needs to be estimated separately \cite{hasan2018downlink, gulati2013learning}. 
Another issue is to design and implement an algorithm to select the optimal antenna modes to maximize the resources available in multiple antenna channels \cite{hasan2018downlink,cetiner2004multifunctional}.
The problem of designing an RA MIMO system, which includes channel estimation, mode selection, and precoder design is a combinatorial optimization problem that is typically NP-hard.
These challenges become more difficult to handle when the dynamic nature of the wireless channel is considered in the modeling of the system. This makes the optimal mode selection of reconfigurable antennas a highly challenging task. 

\subsection{Related Works}
The antenna configuration selection, proposed in \cite{piazza2009technique}, aims to increase the spectral efficiency of the MIMO communication system. The RA mode selection algorithm in \cite{piazza2009technique} exploits spatial correlation of wireless channel to solve the optimization problem.    
In \cite{bahceci2016efficient}, a block minimum mean squared error (MMSE) channel estimation of a multifunctional RA-MIMO system is proposed. The proposed estimation/prediction scheme selects a subset of antenna modes based on an optimal method, and the channel is trained for the chosen modes. Also, based on the correlation between antenna patterns, the channel for the remaining modes is predicted to reduce the channel estimation overhead.

A downlink MU-MIMO transmission for RA systems with an iterative mode selection scheme is developed in \cite{hasan2018downlink}. The proposed low-complexity selection algorithm can operate in real-time. The proposed system assumes perfect channel state information (CSI) at the transmitter, and it applies zero-forcing (ZF) precoding to mitigate the inter-user interference. Multi-objective genetic algorithm (GA) optimization is employed in \cite{hasan2018downlink} to reduce the cardinality of the set of operational modes. 

In \cite{li2020analog}, a hybrid precoder based on an RA system is proposed. The highly reconfigurable antennas (HRA) are the pixel RA antennas that can provide billions of different radiation patterns. Using HRA relaxes the need for phase shifters, and combiners. In \cite{li2020analog}, radiation patterns are expressed via a multi-dimensional vector, and each element of this vector is selected from a set of basis signals based on angle-shifted periodic sinc functions. 
In \cite{bouida2015reconfigurable}, considering correlated, and non-identically distributed Rician fading channels, a mode selection algorithm is proposed that jointly optimizes the fading and polarization correlation parameters to enhance the bit error rate (BER) performance of the system. A low-complexity mode selection scheme is proposed in \cite{hossain2017parasitic} to find the optimal RA mode in response to changes in the channel and user density. An RA architecture in \cite{towfiq2018reconfigurable} is proposed with a control algorithm enabling one to choose the optimal modes of operation. Moreover, a proportional fair scheduler is used for user scheduling, and the optimal mode is selected based on maximizing the weighted sum rate.  

\subsection{Contributions and Novelties}
The contribution of this paper can be stated as follows: 
\begin{enumerate*}[label=(\roman*)]
    \item Introducing the system model of an MU-MIMO system with RAs and formulating the optimization problem for precoder and mode selection design,
    \item Proposing an iterative mode selection algorithm with low computational complexity for data transmission, 
    \item Presenting an offline sectorized mode selection to improve the channel estimation quality, and 
    \item Developing a channel estimation scheme for RA systems for a subset of modes and predicting the channel of remaining modes based on pattern correlation of untrained modes and trained modes.    
\end{enumerate*}

\subsection{Outline}
The paper is organized as follows. In Section \ref{sec:problemForm}, the system model of an RA-based MU-MIMO system is presented. Correspondingly, the joint problem formulation of the precoder and mode selection scheme design for data transmission is provided in Section \ref{sec:problemForm}. The low-complexity iterative mode selection algorithms are introduced in Section \ref{sec:modeSel}. The channel estimation scheme for the RA-MIMO system and the mode selection for channel training is presented in Section \ref{sec:channelEst} and Section \ref{sec:JointChanEstModeSel}, respectively. Section \ref{sec:SimulRes} presents the simulation results. Finally, Section VII concludes the paper.

\section{Problem Statement}
\label{sec:problemForm}
\begin{figure}
\centering     
\subfigure[]{\includegraphics[width=0.9\linewidth]{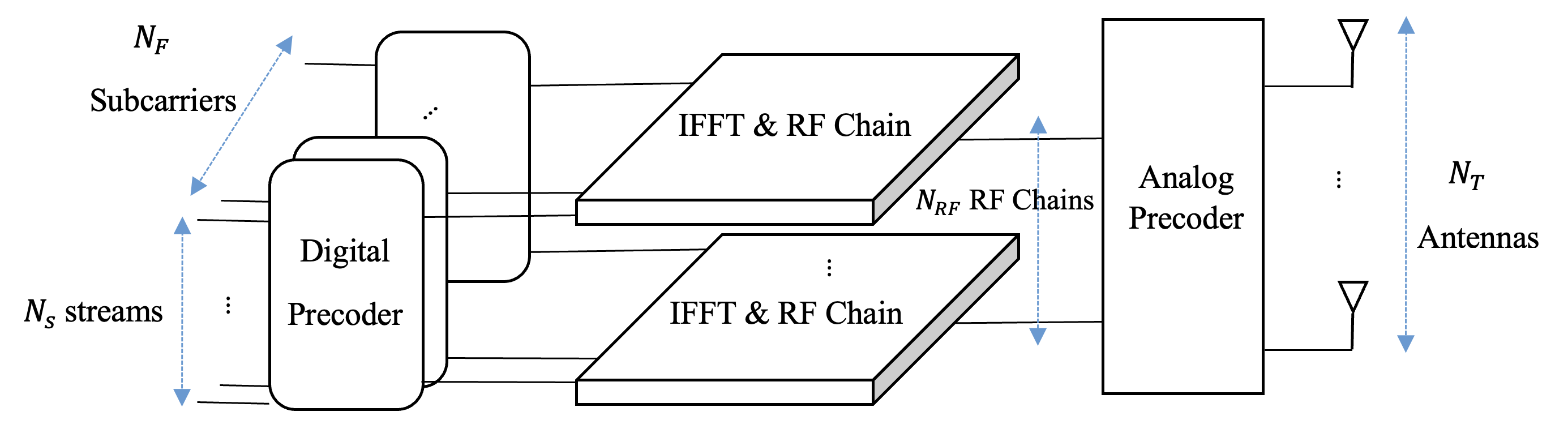}}
\subfigure[]{\includegraphics[width=0.9\linewidth]{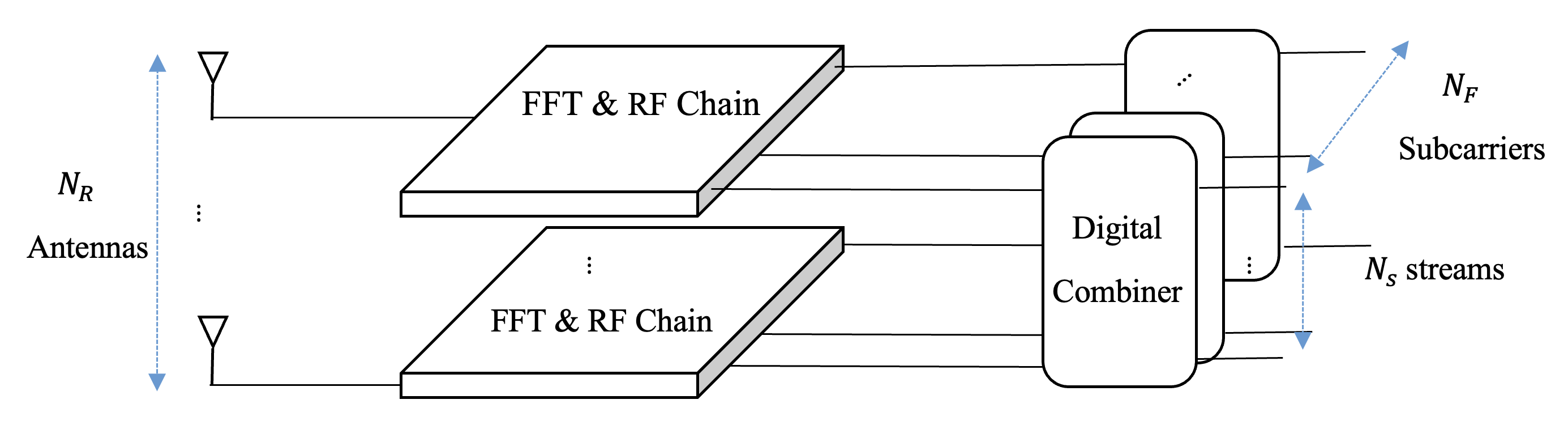}}
\caption{(a) BS and (b) UE, signal transmitter and receiver, respectively, architectures.}
\label{fig.generalarchitecture}
\end{figure}

We consider the downlink of an OFDM-based MIMO wireless communication system with a hybrid beamforming structure. The base station (BS) is equipped with a uniform planar array with $N_T$ antennas and $N_{RF}$ RF chains to simultaneously transmit $N_s$ data streams to the users ($n_s$ data streams per user). We assume $K$ users are scheduled to be served where each user is equipped with $N_R$ antennas. In this paper, RAs are employed at the BS. In particular, antennas with radiation pattern reconfigurability. 
A total of $L=|\mathcal{M}|$ different antenna modes exist, and the operating mode of each $n_t$-th antenna is selected from the pool set $\mathcal{M}$ of radiation patterns.
Fig. \ref{fig.generalarchitecture} represents the architectures of the BS and user equipment (UE), which in the downlink scenario are transmitter and receiver, respectively.

At the BS, the $n_s$ data symbols of the $k$th user $\boldsymbol{s}_k[f,t]$, $k\in\{1,\cdots,K\}$, at time $t$ and at each OFDM subcarrier $f\in\{1,\cdots,N_f\}$ are first precoded using a digital precoding matrix $\boldsymbol{F}_{BB,k}[f,t]\in \mathbb{C}^{N_{RF}\times n_s}$. Concatenating the data symbols and digital precoding matrices of all users results in $\boldsymbol{s}[f,t]\in\mathbb{C}^{N_s\times 1}$ and $\boldsymbol{F}_{BB}[f,t]\in \mathbb{C}^{N_{RF}\times N_s}$. The symbol blocks are transformed to the time domain using $N_{RF}$ $N_f$-point IFFTs. Next, the $N_T\times N_{RF}$ precoder $\boldsymbol{F}_{RF}$ is applied, which is the same for all subcarriers. Here, the RF precoder $\boldsymbol{F}_{RF}$ is assumed to be fixed without phase shifters. As a result, the transmit signal $\boldsymbol{x}[f,t]\in\mathbb{C}^{N_{T}\times 1}$ is given by
\begin{align}
    \boldsymbol{x}[f,t] &= \boldsymbol{F}_{RF}[t] \sum_{\bar{k}=1}^K{\boldsymbol{F}_{BB,\bar{k}}[f,t]\boldsymbol{s}_{\bar{k}}[f,t]}.
\end{align}
The mode vectors of all $N_T$ antennas are represented as $\boldsymbol \mu=[\mu_1,\cdots,\mu_{N_T}]^T\in\mathcal{M}^{N_T\times 1}$, where the mode of the $n_t$-th antenna, $n_t\in\{1,\cdots,N_T\}$, is denoted by $\mu_{n_t}\in\mathcal{M}$. The channel matrix of user $k$ is denoted as $\bH_k[f,t,\boldsymbol \mu] = [\bh_{k,1}[f,t,\mu_{1}],\cdots,\bh_{k,N_T}[f,t,\mu_{N_T}]] \in \mathbb{C}^{N_R\times N_T}$. Subsequently, the network channel of all the users is given by:
\begin{equation}
    \bH(f,t,\boldsymbol \mu) = [\bH_1(f,t,\boldsymbol \mu)^T,\cdots,\bH_K(f,t,\boldsymbol \mu)^T]^T \in \mathbb{C}^{KN_R\times N_T}
\end{equation}
The received signal at the $k$-th user can be represented as
\begin{align}
\boldsymbol{y}_k[f,t,\boldsymbol\mu] &=  {\boldsymbol{H}}_k[f,t,\boldsymbol\mu]{\boldsymbol{F}}_{\text{RF}}[t]{\boldsymbol{F}}_{\text{BB},k}[f,t]{\boldsymbol s_k}[f,t]\notag \\
    &+ \underbrace{{\boldsymbol{H}}_k[f,t,\boldsymbol\mu]{\boldsymbol{F}}_{\text{RF}}[t]\sum_{\bar{k}=1,\bar{k}\neq k}^K {\boldsymbol{F}}_{\text{BB},\bar{k}}[f,t] {\boldsymbol s_{\bar{k}}}[f,t]}_{\text{interference}}\notag \\
    &+ {\boldsymbol z_k}[f,t] ~, ~ k = 1,2,\dots, K
    \label{eq:recSig}
\end{align}
where ${\boldsymbol z_k}[f,t]\sim \mathcal{N}(\mathbf{0},\sigma_n^2\boldsymbol{I})$ is the noise vector at the receiver of user $k$.

Assuming decoupling the joint transmitter and receiver design, the main focus of this paper is on the design of digital precoder ($\boldsymbol{F}_{BB}$) and optimal antenna modes $\boldsymbol \mu$. Therefore, we consider interference as noise, and the achieved sum spectral efficiency will be: 
\begin{equation}
\begin{aligned}
    R = \sum_{k=1}^K\log_2(&|\boldsymbol{I}+\frac{\rho}{N_s}\boldsymbol{C}^{-1}\boldsymbol{H}_k[f,t,\boldsymbol \mu]\boldsymbol{F}_{RF}[t]\boldsymbol{F}_{BB,k}[f,t]\\
    &\boldsymbol{F}_{BB,k}^H[f,t]\boldsymbol{F}_{RF}^H[t]\boldsymbol{H}_k^H[f,t,\boldsymbol \mu]|)
    \label{eq:sumrate}
\end{aligned}
\end{equation}
where $\boldsymbol{C} = {\boldsymbol{H}}_k[f,t,\boldsymbol\mu]{\boldsymbol{F}}_{\text{RF}}[t]\sum_{\bar{k}=1,\bar{k}\neq k}^K {\boldsymbol{F}}_{\text{BB},\bar{k}}[f,t]{\boldsymbol{F}}_{\text{BB},\bar{k}}^H[f,t]$ ${\boldsymbol{F}}_{\text{RF}}^H[t] {\boldsymbol{H}}_k^H[f,t,\boldsymbol\mu] + \sigma_n^2 \boldsymbol{I}$ is the covariance matrix of both interference and noise, and $\rho$ represents the averaged received power.

Proceeding with the design of precoder $\boldsymbol{F}_{BB}$ and mode selection $\boldsymbol \mu$, the most general form of the optimization problem can be stated as
\begin{equation}
\begin{aligned}
    (\boldsymbol{F}_{BB}^{OPT},\boldsymbol \mu^{OPT}) = \underset{\boldsymbol{F}_{BB},\boldsymbol \mu}{\arg \max}~~& {R(\boldsymbol{F}_{BB},\boldsymbol\mu)}\\
    s.t. ~~ & ||\boldsymbol{F}_{RF}\boldsymbol{F}_{BB}||_F^2 = N_s\\
    & \boldsymbol \mu \in\mathcal{M}^{N_T\times 1}
    \end{aligned}
    \label{eq:generaldesign}
\end{equation}
In the spectral efficiency formulation \eqref{eq:sumrate}, we still have the notion of $\boldsymbol{F}_{RF}$, since in general, a one-to-one mapping of RF chains and transmit antennas might not exists (i.e., more than one transmit antenna connected to the same RF chain); however, it is assumed to be fixed and is not an optimization parameter.
To the extent of the author's knowledge, no general solutions to \eqref{eq:generaldesign} are known in the presence of the non-convex feasibility constraint $\boldsymbol \mu \in\mathcal{M}^{N_T\times 1}$. We propose to solve an approximation of \eqref{eq:generaldesign} to find practical near-optimal solutions.
\subsection{Digital Precoder Design}
To design the digital precoder for a multi-user MIMO system, we need to consider \eqref{eq:recSig}. To eliminate the interference term in ${\boldsymbol y_{k}}$, the low-dimensional block diagonalization (BD) processing with digital precoder ${\boldsymbol{F}}_{\text{BB}}$ can be performed \cite{spencer2004zero}. The BD transmission scheme has two stages. In one step, the precoding matrix for each user is designed to suppress the other users' interferences by choosing the null space of other users' matrices as the precoding matrix for the intended user. In the second stage, the precoding matrices should be designed to maximize the system capacity under the zero-interference constraint \cite{spencer2004zero}. 
However, the performance of the BD precoder depends on the availability of perfect CSI at BS, and also it is sensitive to noise. Therefore, the regularized block diagonalization (RBD) algorithm is used as the digital precoder when the CSI is inaccurate at BS. In RBD, the precoding matrix of each user is chosen such that the off-diagonal block matrices of the effective channel converge to zero when SNR increases \cite{stankovic2008generalized}.
\section{RA Mode Selection Scheme Design}
\label{sec:modeSel}

To facilitate the understanding of the structure of $\boldsymbol{H}[\boldsymbol \mu]$, we introduce the notion of a channel candidate pool $\boldsymbol{H}_c[\nu]$, where

\begin{equation}
    \boldsymbol{H}_c[\nu] = \boldsymbol{H}[{\boldsymbol \mu}], \text{ s.t. } \mu_{n_t} = \nu, ~\forall n_t\in\{1,\ldots,N_T\},  
\end{equation}
In other words, $\boldsymbol{H}_c[\nu]$ is the channel when all the transmit antennas select the specific mode $\nu$. Using this new notation and noting that the channel of each transmit antenna (equivalently, the corresponding column of the channel matrix) is selected from one of these candidate channels. Therefore, we denote the selection matrix $\boldsymbol{W}\in\{0,1\}^{N_T\times L}$, in which $W_{n_t,\mu}=1$ means that the $n_t$-th antenna has selected mode $\mu$, and $0$ otherwise. Therefore we can reformulate the channel matrix based on the selection matrix as follows:

\begin{align}
    \boldsymbol{H}[\boldsymbol{W}] &= \sum_{\nu=1}^{L} \boldsymbol{H}_c[\nu] \Diag(\boldsymbol{W}_{\mu})\notag\\
        \text{s.t.} & \sum_{\mu=1}^{L} W_{n_t,\mu} = 1, ~\forall n_t, \notag\\
        & \boldsymbol{W}\in\{0,1\}^{N_T\times L},
        \label{eq.RAMat}
\end{align}
with $\Diag(\cdot)$ denoting the diagonal matrix operator and $\boldsymbol{W}_{\mu}$ is the $\mu$th column of $\boldsymbol{W}$. Note that there is a one-to-one mapping between a $\boldsymbol \mu$ and the corresponding $\boldsymbol{W}$ which results in the same channel matrix, i.e., $\boldsymbol{H}[\boldsymbol \mu] = \boldsymbol{H}[\boldsymbol{W}]$. 

For each RA mode vector $\boldsymbol{\mu}=[\mu_1,\mu_2, \dots, \mu_{N_T}]$, there is a composite MIMO channel $\boldsymbol{H}[\boldsymbol{\mu}]$. Since each of the $N_T$ RAs can switch among $L$ modes, $L^{N_T}$ different MIMO channel states are available. Therefore, an exhaustive search is required to find the optimal set of modes by maximizing the sum spectral efficiency of the MIMO system. However, the exhaustive search through all possible states can be computationally complex for large $L$ or $N_T$. Instead of an exhaustive search, in the following, we develop low-complexity mode selection algorithms to determine the operational RA modes. 

The reduced-complexity heuristic algorithms can exploit different metrics to alternatively optimize RA elements at the BS. The heuristic search reduces the search space at each iteration to $L$ modes, resulting in $LN_T$ steps where all $N_T$ antennas are updated at least once. Therefore, increasing the number of modes or the number of transmit antennas results in a linear increase in the search space compared to the exponential increase in an exhaustive search.
The alternating mode selection algorithm is summarized in Algorithm \ref{alg:Alt-MI} and the intended metric $f(\boldsymbol \mu)$ can be either sum-rate of the system ($f(\boldsymbol \mu) = R(\boldsymbol \mu)$) or summation of channel's eigenvalues ($f(\boldsymbol\mu) =\sum_{i} \lambda_i (\bH[\boldsymbol\mu])$), where $\lambda_i$ is the eigenvalues of the effective channel matrix.

\begin{algorithm}
\caption{Heuristic mode selection algorithm}\label{alg:Alt-MI}
\begin{algorithmic}[1]
\State Initializing modes for transmit antennas for $i=0$ as 
\begin{equation*}
    \bW^{(0)} = \begin{bmatrix}
    1 & 0 & \dots & 0\\
    1 & 0 & \dots & 0\\
    \vdots & \vdots & \ddots & \vdots \\
    1 & 0 & \dots & 0
    \end{bmatrix}
\end{equation*}
\For{$i$-th iteration}
    \For{$n_t=1,\dots,N_T$}
        \State Search through rows of $\bW^{(i)}$ and update $\mu_{n_t}^{\ast}$ such that:
        \begin{align*}
        \mu_{n_t}^{\ast} &= \underset{\boldsymbol{\mu}}{\arg \max}~~ {f(\boldsymbol \mu)}\\
        \text{s.t.} ~& \boldsymbol{\mu} = [\mu_1^{(i)}, \dots, \mu_{n_t-1}^{(i)},\mu_{n_t},\mu_{n_t+1}^{(i-1)},\dots, \mu_{N_T}^{(i-1)}]
    \end{align*}
    \EndFor
    \State $i=i+1$
\EndFor
\label{alg:AltMI}
\end{algorithmic}
\end{algorithm}
\section{RA Systems Channel Estimation}
\label{sec:channelEst}
In Section \ref{sec:modeSel}, all precoders and mode selection schemes are designed while the perfect channel knowledge is assumed. In case the perfect channel knowledge is not available, the impact of the channel measurement errors on the performance of the system should be taken into account. 

In pilot-based channel estimation, the transmitted sequence is assumed to be known on the receiver side. During pilot transmission in the uplink scenario, the $n_t$-th antenna at the BS will receive the following signal at symbol time $t$
\begin{equation}
    y_{n_t}(\mu_{n_t,t})[t,f] = \sum_{n_r=1}^{KN_R} h_{n_t,n_r}(\mu_{n_t,t})[f] x_{n_r}[t,f] + z_{n_t}[t,f].
\end{equation}
The vector $\bmu_{t} = [\mu_{1,t},\mu_{2,t},\dots,\mu_{N_T,t}]$ is the BS antennas' modes at training time $t$. For further simplification, we assume that all antennas at BS select the same mode for training. At the time $t$ the received signal vector at BS can be written as  
\begin{equation}
    \boldsymbol{y}(\boldsymbol\mu_{t})[t,f] = \sum_{n_r=1}^{KN_R} \boldsymbol{h}_{n_r}(\boldsymbol\mu_{t}) x_{n_r}[t,f] + \boldsymbol{z}[t,f]
    \label{eq:ChannelEst}
\end{equation}
where $\boldsymbol{y}(\boldsymbol\mu_{t})[t,f]\in \mathbb{C}^{N_T\times 1}$ and $\boldsymbol{z}[t,f]\in \mathbb{C}^{N_T\times 1}$ are the received signal and the received noise vector, with $E\{z_{n_t}[t,f]\} = 0$ and $E\{z_{n_t}[t,f]z_{n_t}^{\ast}[t,f]\} = \sigma_{z}^2$, at time $t$ and subcarrier $f$, respectively. Moreover, $x_{n_r}[t,f]$ is the pilot symbol of the $n_r$-th UE at period $t$ and frequency tone $f$ with $E\{x_{n_r}[t,f]\} = 0$ and $E\{x_{n_r}[t,f]x_{n_r}^{\ast}[t,f]\} = \sigma_{x}^2$, and $\boldsymbol{h}_{n_r}(\boldsymbol\mu_{t})\in \mathbb{C}^{N_T\times 1}$ is the channel between the $n_r$-th antenna at UE and the BS antennas.
The MMSE channel estimation can be obtained as
\begin{equation}
    \hat{\boldsymbol{h}}_{n_r,\text{MMSE}}(\boldsymbol\mu_t) = \boldsymbol{R}_{{\boldsymbol{h}}_{n_r}{\boldsymbol{h}}_{n_r}}\big(\boldsymbol{R}_{{\boldsymbol{h}}_{n_r}{\boldsymbol{h}}_{n_r}} + \frac{\sigma_z^2}{\sigma_x^2}\bI\big)^{-1}\Tilde{{\boldsymbol{h}}}_{n_r}(\boldsymbol\mu_t)
\end{equation}
where $\Tilde{{\boldsymbol{h}}}_{n_r}(\boldsymbol\mu_t)= \frac{\boldsymbol{y}(\boldsymbol\mu_{t})}{x_{n_r}}$, and auto-correlation $\boldsymbol{R}_{{\boldsymbol{h}}_{n_r}{\boldsymbol{h}}_{n_r}} = E\{{\boldsymbol{h}}_{n_r}(\boldsymbol\mu_t){\boldsymbol{h}}_{n_r}^H(\boldsymbol\mu_t)\}$. 

\subsection{Joint Channel Estimation and Prediction based on Channel Correlation}
It can be seen in \eqref{eq:ChannelEst} that changing the BS antenna modes results in different channel realizations for the same propagation channel. Since it can be infeasible to train and estimate the channel for all antenna modes within the coherence time, an efficient MIMO channel estimation procedure for RA systems should be exploited. Therefore, developing a framework for channel estimation and prediction is required. A small set of antenna modes is selected to be trained, and the channel of the remaining modes is predicted based on the correlation among them. Assuming $F$ antenna modes are selected to train the channel for $n_t$-th antenna at BS and $n_r$-th antenna at UE, the channel for point-to-point communication is presented as $ \boldsymbol{h}_{n_t,n_r} = [h_{n_t,n_r}(\mu_1),\dots,h_{n_t,n_r}(\mu_F)]^T$. The prediction of channel ($\hat{\boldsymbol{h}}_{n_t,n_r} ^c$) for the remaining $L-F$ modes is obtained by MMSE prediction \cite{bahceci2016efficient}:
\begin{equation}
    \hat{\boldsymbol{h}}_{n_t,n_r,\text{MMSE}}^c = \bR_{\boldsymbol{h}^c\boldsymbol{h}}\bR_{\boldsymbol{h}\boldsymbol{h}}^{-1}\hat{\boldsymbol{h}}_{n_t,n_r} = \boldsymbol{R}_{\boldsymbol{h}^c\boldsymbol{h}}\big(\boldsymbol{R}_{\boldsymbol{h}\boldsymbol{h}} + \frac{\sigma_z^2}{\sigma_x^2}\bI\big)^{-1}\Tilde{\boldsymbol{h}}_{n_t,n_r}
    \label{eq: ChanPredCorr}
\end{equation}
where $\boldsymbol{R}_{\boldsymbol{h}^c\boldsymbol{h}}$ is the $F\times(L-F)$ cross-correlation matrix among the channel realizations of $F$ trained modes and the remaining ones ($L-F$), and $\hat{\boldsymbol{h}}_{n_t,n_r}$ is the estimated channel for $F$ modes. The overall estimated and predicted channel for $n_t$-th antenna at BS and $n_r$-th antenna at UE, including all available modes, is shown as $\hat{\boldsymbol{h}}_{\mu}\in \mathbb{C}^{L\times 1}$. 
\subsection{Joint Channel Estimation and Prediction based on Pattern Correlation}
The radiation channel consists of the wireless channel (affected by the propagation environment) and antenna radiation pattern effects. Therefore, the channel gains for different modes differ in their radiation patterns. Since there is a limitation to estimating the channel of all available modes, one can exploit the fact that the correlation between different mode channel realizations depends on the radiation patterns correlations. 

To that end, as long as the propagation medium remains the same, the channel gains for a set of modes can be predicted via their pattern correlations with other sets of trained modes. Therefore, in \eqref{eq: ChanPredCorr}, the channel correlation can be substituted by radiation pattern correlations.
\begin{equation}
    \hat{\boldsymbol{h}}_{n_t,n_r,\text{MMSE}}^c = \boldsymbol{R}_{\boldsymbol{p}^c\boldsymbol{p}}\big(\boldsymbol{R}_{\boldsymbol{p}\boldsymbol{p}} + \frac{\sigma_z^2}{\sigma_x^2}\bI\big)^{-1}\Tilde{\boldsymbol{h}}_{n_t,n_r}
\end{equation}
where $\boldsymbol{R}_{\boldsymbol{p}^c\boldsymbol{p}}$ is the cross-correlation matrix between patterns of untrained modes and trained modes, and $\boldsymbol{R}_{\boldsymbol{p}\boldsymbol{p}}$ is the pattern correlation matrix of trained modes. Knowing that one of the challenging tasks in channel estimation and prediction is the calculation of the channel correlation matrix, obtaining the pattern correlation matrix reduces the complexity significantly.
\section{Channel Estimation and Mode Selection}
\label{sec:JointChanEstModeSel}
In order to select $F$ modes for channel estimation, the optimization problem is defined as below
\begin{equation}
\begin{aligned}
    \boldsymbol\mu^{OPT} = \underset{\boldsymbol \mu}{\arg \min}~~& {||{\boldsymbol{h}}_{\mu}- \hat{\boldsymbol{h}}_{\mu}||^2_2}\\
    s.t. ~~ & \boldsymbol \mu = [\mu_1,\ldots,\mu_F],\\  &\mu_f\in\mathcal{M}, \quad \forall f\in\{1,\ldots,F\},
    \end{aligned}
    \label{eq:ChanEstOptimization}
\end{equation}
where the vector $\boldsymbol\mu$ contains selected modes. To find the optimal training set of modes, the receiver estimates the channel for all possible ${L \choose F}$ combination of modes and select the set with minimum channel estimation/prediction mean squared error (MSE). Then, after channel training for the set of selected modes, the channel is predicted for the remaining modes. Next, the receiver reports the optimal mode for the data transmission for the rest of the coherence time. Therefore, it is vital to determine a suitable value for the number of selected training modes $F$ to prevent channel training overhead. Regarding this concern, there is a trade-off between the data rate of the system and channel estimation performance. For a small $F$, performance degradation in channel estimation is inevitable. On the other hand, for a large number of $F$, there would be fewer resources for data transmission. 

\subsection{Offline Mode Selection Scheme}
Due to the high complexity of the exhaustive mode selection scheme, we propose an offline mode selection scheme. Before the real-time channel training, the modes with minimum MSE are selected. However, the azimuth plane is divided into $N_{\text{sec}}$ sectors to achieve the best performance, and the optimal training set is determined for each sector. Based on the angle of arrival of each user related to each sector, the optimal set of modes is selected for the upcoming channel training periods.
\section{Simulation Results}
\label{sec:SimulRes}
We employ the system model described in Section \ref{sec:problemForm} to evaluate the performance of MIMO systems with RA. The Quadriga toolbox generates the channel in MATLAB \cite{jaeckel2014quadriga}. Stationary users are assumed in random locations between $50$ and $100$ m with arbitrary azimuth angles between $-60^\circ$ and $60^\circ$ and elevation angles between $-15^\circ$ and $15^\circ$. The users' scenario and propagation conditions are specified in the Quadriga toolbox as terrestrial urban macro-cell parameters extracted from measurements in Berlin, Germany, with line of sight (LOS) conditions. The users are equipped with two uniform linear arrays (ULA) antenna elements. The simulation results contain two parts of simulations: one with perfect channel assumption and one for RA system channel estimation.
\subsection{Data Transmission with Perfect Channel}
\begin{figure}
        \centering
        \includegraphics[width=0.49\textwidth]{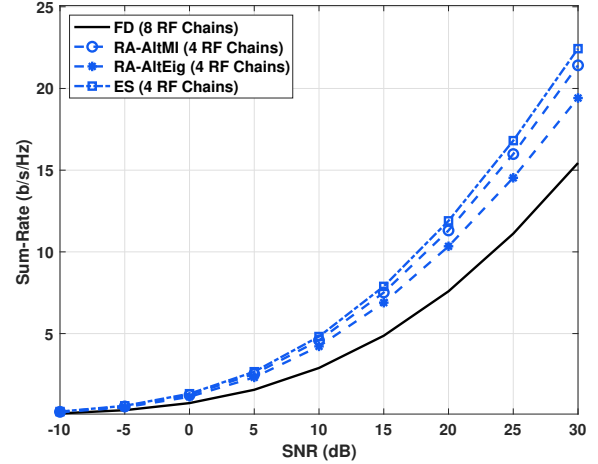}
        \caption{Sum-rate of system (b/s/Hz) versus SNR (dB) in a $2$-user MIMO system with $8$ antennas at BS}
        \label{fig:SmallPanel4RFChain}
\end{figure}

To each user at the receiver side, two data streams are transmitted. In the RA-MIMO systems, $L = 10$ patterns are used for the BS antennas. All simulations in this report were performed for the narrow-band system with one subcarrier. To compare the performance of the RA-MIMO system with digital beamforming (FD), we use the BD algorithm as the digital precoder in a conventional non-RA system. In the FD system, the BS is a panel with $8$ antennas connected to $8$ RF chains, and in the RA-MIMO system, the BS panel contains $8$ RA elements connected to $4$ RF chains. The results are shown in Fig. \ref{fig:SmallPanel4RFChain}.
In Fig. \ref{fig:SmallPanel4RFChain}, "RA-AlMI" and "RA-AltEig" are referred to as two alternating mode selection algorithms with metrics of sum-rate and summation of eigenvalues, respectively. Furthermore, "ES" represents the exhaustive search for selecting the optimal modes for data transmission.  
It is seen that the RA-based MU-MIMO can outperform the conventional antenna-based FD beamforming with more RF chains. In Fig. \ref{fig:SmallPanel4RFChain}, the performance of the proposed heuristic mode selection schemes is depicted. The "RA-AltMI" scheme has a performance close to the optimum method, exhaustive search, while the complexity of "RA-AltMI" is reduced. 
\subsection{Channel Estimation in RA-MIMO System}
\begin{figure}
    \centering
    \includegraphics[width=0.49\textwidth]{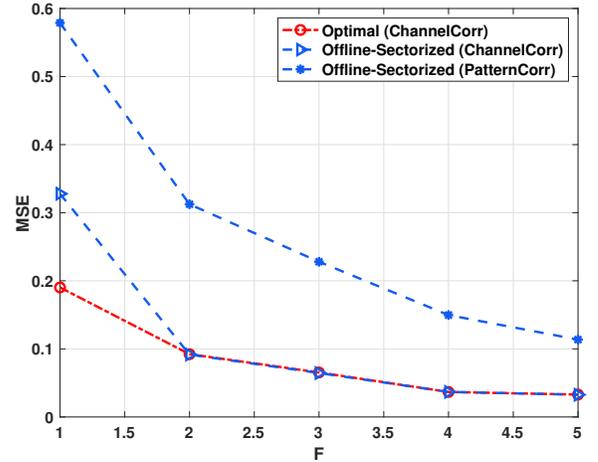}
    \caption{Channel estimation MSE versus the number of selected modes for training ($F$) at $\text{SNR}=20$ (dB)}
    \label{fig:MSEAndF}
\end{figure}

Fig. \ref{fig:MSEAndF} illustrates the normalized MSE of channel estimation/prediction versus the number of selected modes for channel training ($F$) at $\text{SNR} = 20$ dB. As expected, increasing the number of training modes improves the channel estimation/prediction accuracy, however, at the cost of higher channel training overhead.
In the following, all simulations are performed assuming $F = 3$.

\begin{figure}
    \centering
    \includegraphics[width=0.49\textwidth]{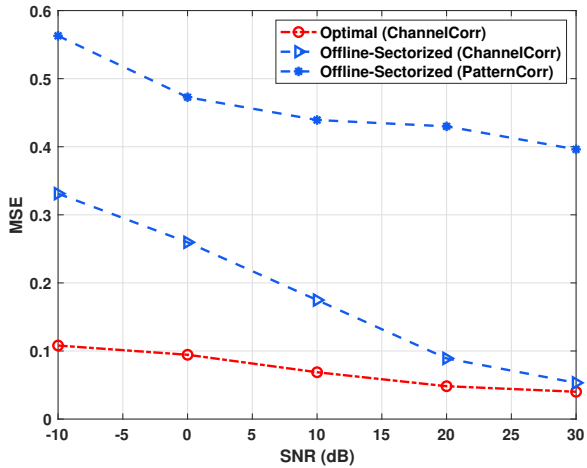}
    \caption{Channel estimation MSE versus SNR (dB) for $F=3$}
    \label{fig:MSEandSNR}
\end{figure}

\begin{figure}
    \centering
    \includegraphics[width=0.49\textwidth]{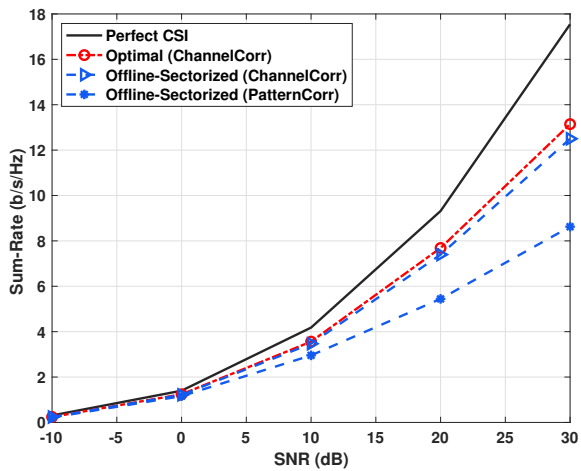}
    \caption{Sum-rate of system (b/s/Hz) versus SNR (dB) for $F=3$}
    \label{fig:capVersusSNR}
\end{figure}

The channel estimation MSE versus SNR is shown in Fig.\ref{fig:MSEandSNR}. In Fig.\ref{fig:MSEandSNR}, the performance of channel estimation based on MSE with two training mode selection methods, optimal (exhaustive search over all possible combinations), and offline modes selection (with $N_{\text{sec}} = 4$), are compared. In the case of offline modes selection, the channel state prediction is performed by channel correlation and pattern correlation among modes. There is a performance loss in channel estimation/prediction based on pattern correlation. However, using pattern correlation in channel estimation/prediction exhibits lower complexity than channel correlation calculation.
The results of system throughput versus the SNR are depicted in Fig. \ref{fig:capVersusSNR}. In Fig. \ref{fig:capVersusSNR}, the sum rate has been calculated when the perfect CSI is available in the RA-MIMO system, and the system's throughput is compared with the system using the estimated channel. As shown in Fig. \ref{fig:capVersusSNR}, there is a performance loss when the estimated channel is used for precoder design. However, to reduce the complexity of the training mode selection algorithm, we use the offline mode selection method. The spectral efficiency of the offline method is reduced slightly compared with the optimal training mode selection method, while the complexity of the searching algorithm is reduced. Furthermore, using pattern correlation for channel prediction of untrained modes results in performance degradation, specifically at higher SNRs.
\section{Conclusion}
\label{sec:Concolusion}
Investigating the benefits of RA systems, in this paper, we tackle two major challenges, namely the combinatorial mode selection optimization problem and the significant channel estimation overhead for all modes. We propose an iterative mode selection algorithm with low computational complexity. Simulation results illustrated near-to-optimal performance (solution of exhaustive search) for small-scale scenarios. Addressing the channel estimation problem, considering a combination of channel estimation and prediction, we propose a novel offline mode selection to perform the channel estimation and subsequently predict the channels of the rest of the modes. Finally, a joint channel estimation and mode selection scheme is demonstrated.

\ifCLASSOPTIONcaptionsoff
\newpage
\fi

\bibliographystyle{IEEEtran}
\bibliography{references}
\end{document}